\documentclass{PoS}

\def\solar{\ifmmode_{\mathord\odot}\else$_{\mathord\odot}$\fi~}

\title{Towards millimeter VLBI}

\ShortTitle{Towards mm-VLBI}

\author{\speaker{T.P.~Krichbaum}, U.~Bach, D.A.~Graham, W.~Alef, A.~Roy, A.~Witzel, J.A.~Zensus\\
        Max-Planck-Institute f\"ur Radioastronomie, Bonn, Germany\\
        E-mail: \email{tkrichbaum@mpifr.de, ubach@mpifr.de, dgraham@mpifr.de, walef@mpifr.de, aroy@mpifr.de, awitzel@mpifr.de, azensus@mpifr.de}}

\author{M. Bremer\\
        Institut de Radio Astronomie Millim\'etrique, Grenoble, France\\
        E-mail: \email{bremer@iram.fr}}

\author{S. Sanchez\\
        Instituto de Radioastronomia Millimetrica, Granada, Spain\\
        E-mail: \email{sanchez@iram.es}}

\abstract{We discuss the present performance and the future perspectives of
VLBI in the 3\,mm to 0.85\,mm observing bands (so called mm-VLBI).
The availability of new telescopes for mm-VLBI and the recent
technical development towards larger observing bandwidth and higher data-rates 
now allow to image with 3mm-VLBI hundreds of sources with high dynamic range.
At 1.3\,mm, pilot VLBI studies have proven detectability of the brightest AGN,
and the existence of ultra-compact regions therein. 
In the next few years global VLBI imaging will be established also at 1.3\,mm
wavelength. With an angular resolution in the $10-20$\,$\mu$ mas range, future 
1.3\,mm- and 0.8\,mm VLBI will be an extraordinarily powerful astronomical
observing method, allowing to image the enigmatic `central engines' and the 
foot-points of AGN-jets in greater 
detail than ever possible before. A sufficiently large number of telescopes 
is a prerequisite for global aperture synthesis imaging. Therefore 
a strong effort is needed to make more telescopes available for VLBI at short millimeter
and sub-millimeter wavelengths. In this context, the further VLBI upgrade of both IRAM telescopes and 
the outfit of the APEX telescope in Chile in preparation for later 
mm-/sub-mm VLBI with ALMA is of high scientific importance. With a sufficiently large
mm-VLBI network, the micro-arcsecond scale imaging of the post-Newtonian emission zone around the event horizon/ergosphere of nearby super-massive Black Holes (such as e.g. Sgr\,A*, M\,87, ...) 
should become possible within the next few years.
}

\FullConference{The 9th European VLBI Network Symposium on The role of VLBI in the Golden Age for Radio Astronomy and EVN Users Meeting\\
		 September 23-26, 2008\\
		 Bologna, Italy}

\begin{document}

\section{Introduction}
The physical processes of energy release and the details of the jet 
launching, particle and jet acceleration in the nuclei of active galaxies (AGN) is still
enigmatic and leads to many questions not yet answered. One way to address this
problem, is the direct imaging of the central regions of these objects, with the
highest possible angular and spatial resolution, 
and at wavelengths short enough to look through
existing source intrinsic opacity barriers. The angular resolution
of an interferometer is proportional to the observing wavelength ($\lambda$)
and inversely proportional to the separation ($D$) between its outer elements. 
This leads to millimeter VLBI (mm-VLBI), which observes at the shortest 
possible wavelength using telescopes on ground ($D <$ Earth diameter),
and to space-VLBI 
utilizing baselines larger than the Earth's diameter through an orbiting 
radio antenna in space (VSOP2, RADIOASTRON). The planned space-VLBI missions
will give an angular resolution of up to $\sim 40$\,$\mu$as at 43\,GHz for VSOP2 
and up to $\sim 10$\,$\mu$as at 22\,GHz for RADIOASTRON. These numbers compare well
with the angular resolution of ground based mm-VLBI: $\sim 45$\,$\mu$as for
present day global VLBI at 86\,GHz (GMVA), and  $\sim 17$\,$\mu$as, respectively
$\sim 11$\,$\mu$as for future VLBI at 230\,GHz and 345\,GHz. Owing to its
larger collecting area and observing bandwidth, it is likely that future
mm-/sub-mm VLBI will be more sensitive than the two planned
space-VLBI missions (limitations of space-antenna size and down-link data
rates), allowing to study more objects. Because of their different 
frequency coverage and  sensitivity space-VLBI and mm-VLBI are very complementary.
Both can reach the smallest spatial scales, however at very different 
frequencies and opacities, facilitating the detailed study of the spectral 
characteristics of the corresponding emission regions.
In the more distant future, these regions may be also investigated 
with near-infrared interferometry at the VLTI (GRAVITY, 2012: 
resolution $\sim 1$\,mas, astrometric accuracy $\sim 10$\,$\mu$as, 
\cite{Eisenhauer08}) and with NASA's X-ray interferometry
mission (MAXIM, 2020: resolution $\sim 100$\,$\mu$as, \cite{Cash05}).
The common driver of all these projects is the desire to see and directly image
the central Black Hole and the post-Newtonian region around it.

%
%
%
%


\section{Present and future 3mm-VLBI}
The global 3mm-VLBI array (GMVA: \underline{G}lobal \underline{M}illimeter \underline{V}LBI \underline{A}rray, http://www.mpifr-bonn.mpg.de/div/vlbi/globalmm) presently consist of
13 radio-telescopes, comprising 5 European stations (Effelsberg: 100\,m, Onsala: 20\,m, Metsahovi
14\,m, Pico Veleta: 30\,m and the 6 x 15\,m Plateau de Bure interferometer) and 8 VLBA
stations (2 VLBA stations (HN, SC) are not equipped with 3mm receivers).
In its standard observing mode, the GMVA records data
at a recording rate of up to 512 Mbit/s (128\,MHz bandwidth) using the Mark5 VLBI 
system. It is expected that the recording rates can be increased to up to 4\,Gbit/s
in the near future ($\geq 2009$).
In optimal weather conditions, the single baseline detection threshold
on the most sensitive baselines to IRAM is $70-100$\,$\mu$Jy ($7\sigma$), and 
on baselines to the VLBA is $100-350$\,mJy.
Under realistic observing conditions, the GMVA offers an array sensitivity of
$\sim (1-3)$\,mJy/hr, which allows to image a compact 300\,mJy radio source with a dynamic
range of up to several hundred, during a 12\,hr full uv-coverage observation.
Mainly due to VLBI scheduling constraints at some observatories, the GMVA presently 
observes only twice per year, in spring and autumn.
These dates result from a compromise between the requirement of 
reasonably good 3mm-VLBI weather at all sites (no rain, sky conditions being sufficient for
pointing and calibration measurements at the large telescopes, $\tau \leq 0.2$), 
and the need to allocate the good winter-time for single 
dish mm-/sub-mm observations at IRAM. Over the last years, however, the 
frequency and VLBI agility of the participating 
GMVA stations has improved, so that there should be now no other limitation 
for more frequent VLBI observing than mutual agreement between observatories. 
More frequent global 3mm-VLBI observations are very desirable in view of the known rapid
source and jet kinematics, with apparent jet speeds of up to $\sim 30$\,c and 
angular separation rates of up to $\sim 10$ beams per year 
(GMVA observing beam: 0.04 - 0.06 $\mu$as). The GMVA is open to all
users of the scientific community, calling for proposals twice per year. 
The proposal deadlines are synchronized with those of the NRAO and the EVN. 
Since its establishment as successor of the former Coordinated Millimeter VLBI Array (CMVA),
which previously was organized by the MIT-Haystack Observatory, the
GMVA has observed since April 2004 about 73 projects in 10 observing sessions, 
from scientist of many different countries. We note that the GMVA is operated
on a best effort basis by the staff of the participating observatories and of the 
VLBI correlator of the MPIfR, without a dedicated financial budget assigned 
for the operation of such a global mm-VLBI network. 

The sensitivity and performance of a VLBI network mainly depends on
observing bandwidth, antenna collecting area and uv-coverage. The ongoing development
of the VLBI data acquisition and recording system is now in a transition from
previously analog and narrow bandwidth, to future digital and broad-band recording. 
It is expected that the new digital recording systems (DBE/DBBC) will
become available for the EVN and the VLBA in 2009/2010. The resulting
increase of the observing bandwidth from 128\,MHz to
up to 1\,GHz, which is accompanied by the capability of the new Mark5C recording system to
record with up to 4 Gbit/s data rate, will lead to a factor of 2.8 sensitivity
improvement. This will allow to observe hundreds of new
radio sources (with correlated flux densities $S_{\rm 86 GHz}\geq 10-20$\,mJy). A further increase
of the observing bandwidth is highly desirable (8-16 GBit/s already being discussed),
however, is restricted by the limited bandwidth (500\,MHz) of the current IF distribution 
systems at the VLBA and some European telescopes. The further sensitivity improvement
of mm-VLBI therefore depends also on the inclusion of telescopes with a large 
collecting area. It is obvious that local interferometers such as PdB, CARMA and ALMA will 
play a key role in this. At the shorter wavelength ($\lambda < 3$\,mm), new telescopes,
which are designed for observations in the mm-/sub-mm bands (antenna surface accuracy 
$\sigma \leq \lambda/20$), must be outfitted for VLBI (hydrogen maser, 
VLBI data acquisition, stable LO). To facilitate the future
VLBI recording at high GBit/s rates, all telescopes must have receivers and IF systems,
which are several GHz wide. In the case of the interferometers, such as the IRAM-PdBI,
where the correlator design doesn't support VLBI with rates $> 1$ GBit/s, it may be necessary 
to build so called phased array processors (see below), allowing to phase 
the interferometer for VLBI with a bandwidth of more than 1\,GHz.

The need for additional VLBI stations is not only motivated by the increase of 
total collecting area and sensitivity, but also by a more uniform uv-coverage.
The rapid variability of the atmospheric water vapor and opacity, and the limitations
of the a-priori calibration of antennas (gain) and receivers ($T_{\rm sys}$, $T_{\rm cal}$),
enforce the use of closure amplitudes for the calibration of the visibility amplitudes.
A relatively large number of telescopes ($N > 10-12$) is required to 
facilitate a good convergence of the amplitude self-calibration in all regions 
of the uv-plane, at short and at long uv-spacings. 

For global VLBI in the 3mm band (80-95 GHz), the following other/new telescopes are 
well suited and one can hope that they will be equipped for mm-VLBI soon: 
2x VLBA (HN, SC), Yebes (Spain), Noto (Italy), GBT (VA, USA), CARMA (CA, USA), LMT (Mexico), 
SRT (Sardinia, Italy), MOPRA (Australia), and ALMA (Chile). The participation of the 40\,m Yebes 
antenna in 3mm-VLBI is forseen for the very near future (3\,mm VLBI tests planned for 2009). 
In Figure 1 (left) the expected improvement of the uv-coverage is shown. 
We note the new short baselines between Yebes and
the two IRAM telescopes (Yb-PV: 384\,km, Yb-PdB: 867\,km), which will be extremely
useful for a better determination of the total flux in the VLBI maps and for 
the imaging of the partially resolved mas-scale structures in AGN-jets. 
With the short and sensitive baselines from Yebes to IRAM and to Effelsberg,
the VLBI imaging of the SiO masers in stellar envelopes and of galactic and
extra-galactic absorption line systems will improve substantially.
For some of the other telescopes mentioned above, the time-scale of their 
participation in global 3mm-VLBI is less clear. Some of these stations still
lack VLBI equipment (H-maser, Mark5 recorder), or 3mm receivers
(VLBA(HN), VLBA(SC), Noto, GBT), or phasing capabilities (CARMA),
or are still under construction (LMT, SRT, ALMA). 
We note that the Australian MOPRA station, which is equipped with a 3mm receiver 
and H-maser, would have about 4-6 hrs common visibility with the VLBA for a source like 3C\,273,
and $\sim 9$\,hrs mutual visibility for Sgr\,A* with the VLBA station on Mauna Kea.
 
Owing to their large collecting area the GBT, the SRT, the LMT and CARMA would provide a
substantial improvement of the overall array sensitivity of the GMVA, 
allowing to reach the 50$\mu$Jy level after 12 hrs on source integration 
(full uv-coverage, 4 Gbit/s). Certainly the biggest step in sensitivity would come 
when ALMA joins mm-VLBI with its 50 telescopes, each of 12\,m diameter. 
A technical problem, which needs to be solved, is the phasing of such an array, 
so that it can operate at several GHz bandwidth as a tied array in VLBI mode. Efforts
in this direction are now underway. A phased array processor for 
the SMA is being built \cite{Weintroub08}, which should help to solve this problem 
also for interferometers like PdB, or CARMA. 
In mm-VLBI, the participation of ALMA as a tied array would lower the single 
baseline detection threshold dramatically, in the 3mm band down to a few mJy 
on the most sensitive baselines. This will push global mm-VLBI imaging to 
a similar sensitivity level as seen in present day cm-VLBI, allowing to 
image with micro-arcsecond resolution all compact astronomical objects with
an apparent brightness temperatures as low as $> 10^7- 10^8$\,K.

\begin{figure}
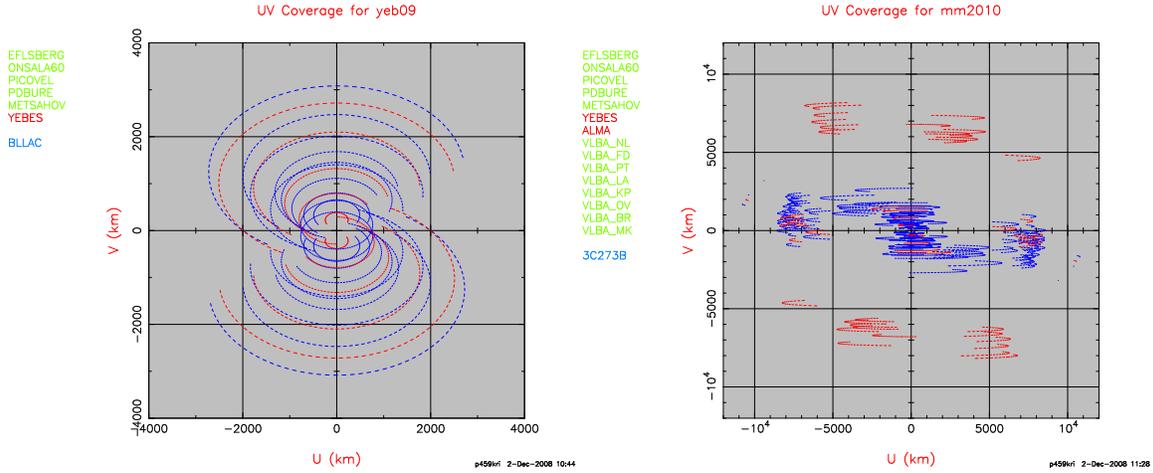

\includegraphics[angle=-90,width=0.5\textwidth]{yeb09UV.cps}
\includegraphics[angle=-90,width=0.5\textwidth]{3mm-3C273-UV.cps}
\caption{{\bf Left:} Simulated uv-coverage for a European 3mm-VLBI experiment, including the
existing European 3mm-VLBI stations (Effelsberg, Onsala, Metsahovi, Pico Veleta, Plateau de Bure)
and the new Yebes telescope (high-lighted in red). 
{\bf Right:} The uv-coverage for a global 3mm-VLBI experiment, including the
existing 3mm-VLBI stations of the GMVA (Europe plus VLBA)
and a VLBI telescope at the ALMA site in Chile. The contribution of Yebes and ALMA
to the uv-coverage are both high-lighted in red, with the following distinction in the 
v-coordinate:  ${\rm v}_{\rm ~Yebes} \leq 3000$\,km, 
${\rm v}_{\rm ~ALMA} \geq 4000$\,km.}
\label{3mmuv}
\vspace*{-0.75cm}
\end{figure}

\section{Highest angular resolution mm-VLBI observations of Cygnus A}
As an example of the capabilities of the GMVA and of global 3mm-VLBI in general, 
we present here some new results from global 3mm-VLBI observations of the two-sided 
jet of Cygnus\,A. Results for other sources have been already presented 
elsewhere (e.g. \cite{Krichbaum06a}, \cite{Agudo07}, \cite{Krichbaum08}, \cite{Giroletti08}).
At a luminosity distance for Cygnus A ($z=0.056075$) of 247\,Mpc,
an angular scale of 0.1\,mas corresponds 
to 0.11\,pc. Full uv-coverage GMVA observations were performed on October 16, 2005 at 86\,GHz
(512 Mbit/s, single polarization), with the participation of 
12 stations (Pico Veleta was weathered out). Cygnus A was detected
at all stations with a correlated flux ranging from 0.7\,Jy at the shortest uv-spacings
to $\sim 80$\,$\mu$Jy at the maximum uv-distance of 3.1\,G$\lambda$ (baselines: PdB-VLBA\_MK, EB-VLBA\_MK). These detections demonstrate the excellent sensitivity of the GMVA,
and in particular of its most sensitive station, the IRAM-interferometer on Plateau de Bure in the French
Alpes. From this experiment a new upper limit to the unresolved VLBI core of $\leq 46$\,$\mu$as
is derived, which corresponds to a spatial scale of $\sim 200$ Schwarzschild radii (assuming
a central BH of $10^9$\,M$\solar$). For a better determination of the jet-to-counter-jet
ratio and for the measurement of the spectral index gradient along the jet, 
uv-tapered images were produced. In Figure 2,
we show two quasi-simultaneously obtained uv-tapered VLBI maps at 43\,GHz (top, Feb. 2005) 
and 86\,GHz (bottom, October 2005). Both maps are convolved with identical beams. The 
maps compare well with each other, both showing emission on the
jet and on the counter-jet side, with the western jet being brighter. In Figure 3
(left), we show the spectral index between 43\,GHz and 86\,GHz ($\alpha_{\rm 43/86\,GHz}$),
which was determined using Gaussian component fits to the data. This
confirms our previous finding from observations at longer wavelength
(\cite{Bach05} and ref. therein) that the eastern counter-jet
exhibits an inverted spectrum, whereas the spectrum of the western jet is optically thin and steep.
In Figure 3 (right), we plot the jet-to-counter-jet ratio as a function of frequency,
adding (for the first time) a measurement of the jet-to-counter-jet ratio also at 86\,GHz
(red symbol). The frequency dependence of the jet-to-counter-jet ratio and the
spectral index distribution (Fig. 3, left) both support the idea of an partially
opaque absorber (e.g. a torus), which covers the inner region of the counter-jet
(but not the jet), and which becomes more transparent towards higher frequencies
(\cite{Kri98}). The new VLBI maps and the spectral index distribution now suggest
that the inner radius of the torus is small, of sub-parsec scale dimensions. The sudden
intensity-drop seen in Figure 2 on the counter-jet side beyond $r \simeq 0.2 - 0.3$\,mas,
and the steepening of the counter-jet spectrum towards the core (from inverted to flat),
also indicates reduced, or even vanishing absorption at small core separations and 
high frequencies. In this case the inner radius of the torus could be of order $\leq 0.3$\,pc.
The interpretation of a $0.2$\,mas emission gap seen in the counter-jet in a recent 
7\,mm VLBI image (October 2007, see \cite{Bach08}) still is unclear and
may indicate motion of a relatively bright counter-jet component or time variability of
the circum nuclear absorber. 

\begin{figure}
\includegraphics[angle=0,width=0.7\textwidth]{comb2.2005.forBologna.ps}
\caption{VLBI image of Cygnus\,A at 43\,GHz, epoch 2005.09 (top) and
86\,GHz, epoch 2005.79 (bottom). Contour levels are at -0.5, 0.5, 1, 2,
4, 8, 16, 32, and 64 \% of the peak flux of 0.26 Jy/beam (top), and 0.25 Jy/beam (bottom).
Both maps are rotated clock-wise by $16^\circ$ and are convolved with the 
same observing beam of 0.20 x 0.10 mas, p.a.=$0^\circ$. 
The lower image shows for the first time the counter-jet of Cygnus\,A also at 86\,GHz.
}
\label{Cygmaps}
~~~\\
\begin{minipage}[t!]{0.49\textwidth}{
\includegraphics[angle=0,width=0.88\textwidth]{spectrum_bologna.eps}
}
\end{minipage}
~~~
\begin{minipage}[t!]{0.49\textwidth}{
\vspace*{2mm}
\includegraphics[angle=0,width=1.02\textwidth]{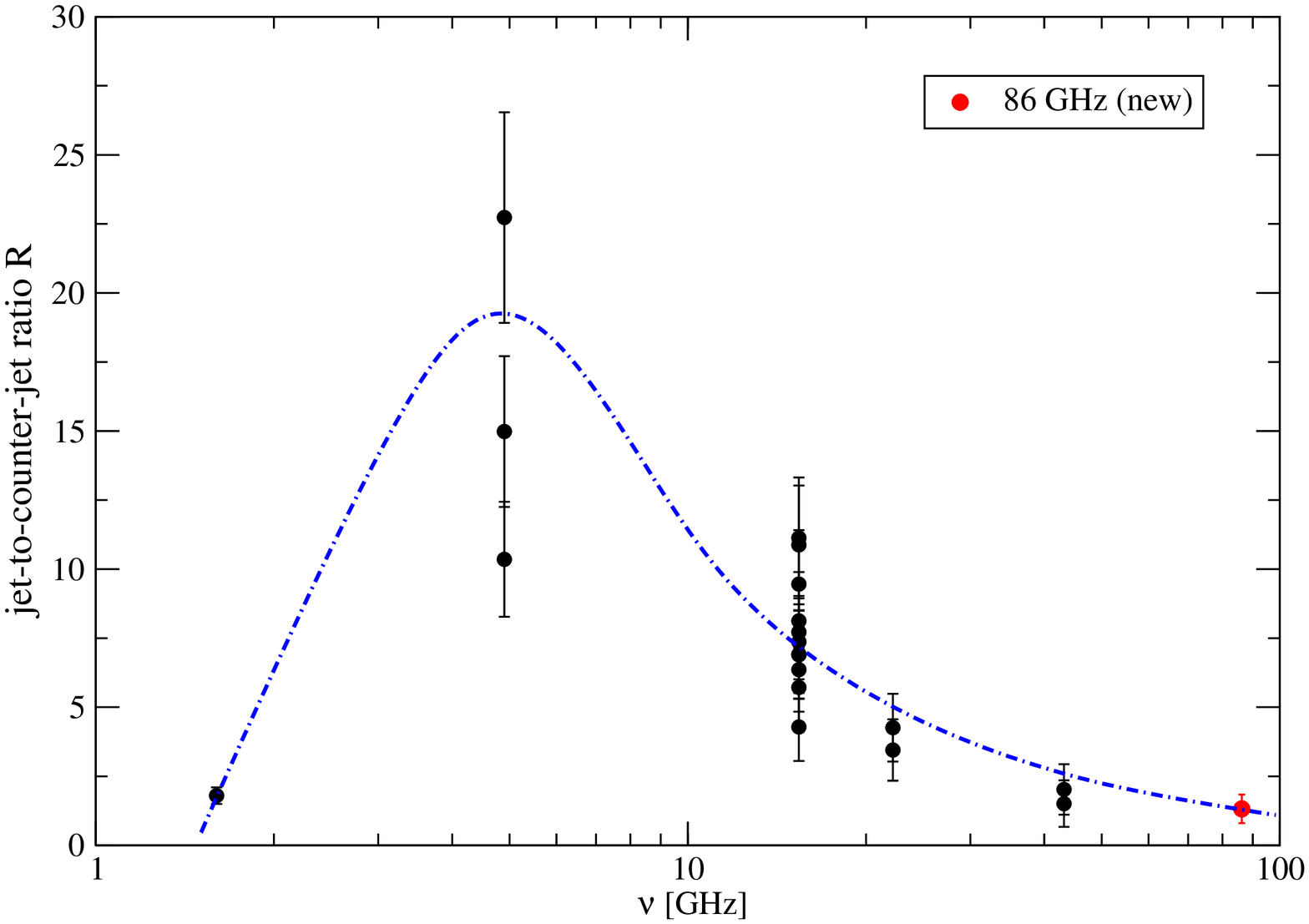}
}
\end{minipage}
\caption{{\bf Left:} The spectral index distribution $\alpha_{\rm 43/86 GHz}$
from counter-jet to jet ($S \propto \nu^\alpha$) plotted versus core-separation
and using the two quasi-simultaneous maps of Figure 2.
{\bf Right:} The jet-to-counter-jet ratio of Cygnus A, plotted versus frequency.
The measurement from the new 3mm-VLBI image of Figure 2 is added in red. Data at
lower frequencies are taken from Bach et al. (2005), and references therein. 
}
\label{cjet}
\end{figure}

\section{Future VLBI at and below 1.3\,mm wavelength}
While VLBI observations at $\lambda=3$\,mm and longer wavelength are now being performed
regularly, all VLBI experiments at shorter wavelength still bear the character of
pilot studies and reflect pioneering work towards future global VLBI at 
$\lambda \leq 1.3$\,mm. Despite many technical difficulties, limitations in the number 
of antennas capable to observe at 1.3\,mm, and a still limited baseline sensitivity, 
some progress has been already made. After the demonstration of technical
feasibility of 1\,mm VLBI in Europe and the verification of sufficiently high 
compactness of several AGN (\cite{Greve95}, \cite{Krichbaum97}) and the Galactic
Center source Sgr\,A* (\cite{Krichbaum98}, \cite{Krichbaum06}), 
first transatlantic fringes between the IRAM 30\,m 
telescope (PV) and the Heinrich Hertz telescope (HHT) on Mt. Graham (Arizona) 
were found at 230\,GHz in 2003 (\cite{Krichbaum04}, \cite{Doeleman05}).
These observations were performed with a data rate of 512\,Mbit/s (at this time still using a prototype
Mark5 system), but suffered from high system temperatures at the HHT (T$_{sys} \sim 1100$\,K)
and some LO instabilities at PdB. Despite these limitations, two sources (3C454.3, 0716+714) were detected with an of $SNR \simeq 7$ on the transatlantic PV-HHT baseline, marking 
a world record fringe spacing of $\sim 6$ G$\lambda$ (angular resolution: $\sim 34 \mu$as). 
The fact that $\sim 50$\,\% of the observed sources (N$=18$) were also clearly 
detected on the 0.9 G$\lambda$ long baseline between PV and PdB (at $SNR=8-25$, \cite{Krichbaum04}), 
outlines the high sensitivity of this baseline and marks 
its importance for future global VLBI at 1.3\,mm wavelength.

It took another 4 years, until April 2007, when a new 1.3\,mm VLBI experiment was
set up, this time with the participation of the HHT (Arizona), a single 10.4\,m antenna
of CARMA (California), the JCMT (Mauna Kea) (fiber linked to the SMA, which provided 
data acquisition and H-maser) and Pico Veleta (Spain). Due to problems with a new
digital VLBI backend (DBE) and partially bad weather, no fringes to PV were found.
With the new DBE, a data rate of 3.84\,Gbit/s (corresponding to a recording bandwidth of
almost 1\,GHz) could be achieved, facilitating the detection of Sgr\,A* (and a number
of other AGN) on the US-American triangle (\cite{Doeleman08}). The resulting new
estimate of the size for the Galactic Centre ($\sim 37$\,$\mu$as, corresponding to $\sim 3.7$
Schwarzschild radii) challenges the theoretical interpretation and of course motivates
the more detailed future imaging. With more VLBI telescopes, the immediate vicinity of the SMBH in Sgr\,A* and
other nearby objects like e.g. M\,87 can be mapped (see \cite{Krichbaum06a}, \cite{Krichbaum08}).
With regard to the 1.3\,mm VLBI observation of 2003, the increase of the observing bandwidth
to $\sim 4$\,Gbit/s, now leads to an improvement in baseline sensitivity
by a factor of $\sim 3$. This and the improved 1.3\,mm receiver performances, now 
should allow to detect with 1.3\,mm VLBI any compact radio source with a correlated flux density
of $\geq 100-200$\,mJy ($7\sigma$, $t_{\rm coh}=10$\,s, 4\,Gbit/s) on IRAM - USA baselines. 
In Table 1 we summarize the characteristics for those radio telescopes, which 
in the next few years are likely to participate
in global VLBI at wavelength of $\leq 1.3$\,mm. It is obvious that
IRAM with its two large mm-telescopes and APEX (run by an MPIfR/OSO/ESO consortium)
located at a favorable southern location in Chile could 
(and should) play an active role in the future development of mm-/sub-mm VLBI.

\begin{table}
\caption{Properties of Radio Telescopes suitable for VLBI at $\nu=230$\,GHz and above}

{\small
\begin{tabular}{lllcccccl}
Station &Location       &Country  &Altitude& D       & Surface&  Eff.    &SEFD$^1$ &H-maser \\
        &               &         & [m]    &   [m]   &[$\mu$m]&          &[Jy]     &      \\
PdB-1   &Plateau de Bure&  France & 2550   &   15    &   55   &  0.45    & 5182    &  yes \\
PdB     &Plateau de Bure&  France & 2550   &  6x15   &   55   &  0.45    &  864    &  yes \\
PV      &Pico Veleta    &  Spain  & 2900   &   30    &   67   &  0.39    & 1485    &  yes \\
APEX    &Chajnantor     &  Chile  & 5100   &   12    &   18   &  0.58    & 6295    &  no  \\
HHT     &Mt. Graham     &  AZ, USA& 3100   &   10    &   15   &  0.59    & 8979    &  yes \\
KP      &Kitt Peak      &  AZ, USA& 2000   &   12    &   75   &  0.35    &10322    &  no  \\
JCMT    &Mauna Kea      &  HI, USA& 4100   &   15    &   25   &  0.57    & 4141    &  yes$^c$ \\
CSO     &Mauna Kea      &  HI, USA& 4100   &   10    &   25   &  0.57    & 8618    &  yes$^c$ \\
SMA     &Mauna Kea      &  HI, USA& 4100   &   8x6   &   12   &  0.59    & 3093    &  yes     \\
Hawai-6 &Mauna Kea      &  HI, USA& 4100   &   $^a$  &   25   &  0.57    & 1696    &  yes$^c$ \\
CARMA-1 &Cedar Flat     &  CA, USA& 2200   &   10.4  &   60   &  0.43    &11373    &  yes     \\
CARMA   &Cedar Flat     &  CA, USA& 2200   &   $^b$  &   60   &  0.43    & 1142    &  yes     \\
LMT     &Sierra Negra   &  Mexico & 4600   &   50    &   70   &  0.38    &  556    &  no \\
ALMA-1  &Chajnantor     &  Chile  & 5000   &   12    &   25   &  0.57    & 6469    &  no \\
ALMA    &Chajnantor     &  Chile  & 5000   &  50x12  &   25   &  0.57    &  129    &  no \\
        &               &         &        &         &        &          &         &     \\  
\end{tabular}
\noindent
\hspace*{2mm} $^1:$ assuming $T_{\rm sys} = 150$ K \\
\hspace*{2mm} $^a:$ the phased Hawaii array consists of the combination of SMA + CSO + JCMT               \\
\hspace*{2mm} $^b:$ 6 x 10.4m + 9 x 6.1m           \\
\hspace*{2mm} $^c:$ via fiber link to SMA          \\
}
\vspace*{-0.5cm}
\end{table}

\end{document}